\documentclass[10pt,sort&compress]{article}
\usepackage{graphicx,rotating,multirow}
\usepackage{bm,multirow}
\usepackage{amsmath,amssymb,color,mathrsfs}

\usepackage{footnote}
\usepackage{citesort}
\usepackage{cuted}

\def\lambdabar{\lambda\kern-1ex\raise0.55ex\hbox{--}}

\newcommand{\APB}{Ann. Phys. (Berlin) }

\newcommand{\CJP}{Can. J. Phys. }

\newcommand{\EJP}{Eur. J. Phys. }

\newcommand{\jpa}{J. Phys. A }

\newcommand{\PLA}{Phys. Lett. A }
\newcommand{\PR}{Phys. Rev. }
\newcommand{\PRA}{Phys. Rev. A }

\newcommand{\PRL}{Phys. Rev. Lett. }

\newcommand{\RMP}{Rev. Mod. Phys. }
\newcommand{\SPJ}{Sov. Phys. - JETP }

\newcommand{\ZETF}{Zh. Eksp. Teor. Fiz. }
\newcommand{\ZP}{Z. Phys. }
\newcommand{\ZPB}{Z. Phys. B }

\definecolor{officegreen}{rgb}{0,0.5,0}
\definecolor{pakistangreen}{rgb}{0,0.4,0}
\definecolor{palatinatepurple}{rgb}{0.41,0.16,0.38}
\definecolor{sangria}{rgb}{0.57,0,0.04}

\begin{document}
\title{Theory of the Robin quantum wall in a linear potential. II. Thermodynamic properties}
\author{O. Olendski\footnote{Department of Applied Physics and Astronomy, University of Sharjah, P.O. Box 27272, Sharjah, United Arab Emirates; E-mail: oolendski@sharjah.ac.ae}}

\maketitle

\begin{abstract}
A theoretical analysis of the thermodynamic properties of the Robin wall characterized by the extrapolation length $\Lambda$ in the electric field $\mathscr{E}$ that pushes the particle to the surface is presented both in the canonical and two grand canonical representations and in the whole range of the Robin distance with the emphasis on its negative values which for the voltage-free configuration support negative-energy bound state. For the canonical ensemble, the heat capacity at $\Lambda<0$ exhibits a nonmonotonic behavior as a function of the temperature $T$ with its pronounced maximum unrestrictedly increasing for the decreasing fields as $\ln^2\mathscr{E}$ and its location being proportional to $(-\ln\mathscr{E})^{-1}$. For the Fermi-Dirac distribution, the specific heat per particle $c_N$ is a nonmonotonic function of the temperature too with the conspicuous extremum being preceded  on the $T$ axis by the plateau whose magnitude at the vanishing $\mathscr{E}$ is defined as $3(N-1)/(2N)k_B$, with $N$ being a number of the particles. The maximum of $c_N$ is the largest for $N=1$ and, similar to the canonical ensemble, grows to infinity as the field goes to zero. For the Bose-Einstein ensemble, a formation of the sharp asymmetric feature on the $c_N$-$T$ dependence with the increase of $N$ is shown to be more prominent at the lower voltages. This cusp-like dependence of the heat capacity on the temperature, which for the infinite number of bosons transforms into the discontinuity of $c_N(T)$, is an indication of the phase transition to the condensate state. Some other physical characteristics such as the critical temperature $T_{cr}$ and ground-level population of the Bose-Einstein condensate are calculated and analyzed as a function of the field and extrapolation length. Qualitative and quantitative explanation of these physical phenomena is based on the variation of the energy spectrum by the electric field.
\end{abstract}

\section{Introduction}\label{Introduction}
Analysis of the interaction between the electric field $\mathscr{E}$  applied perpendicularly to the plane $\cal S$ and  the Robin boundary condition (BC) at it \cite{Gustafson1}
\begin{equation}\label{Robin1}
\left.{\bf n}{\bm\nabla}\Psi\right|_{\cal S}=\left.\frac{1}{\Lambda}\Psi\right|_{\cal S}
\end{equation}
has shown that the spectrum of the one-dimensional quantum particle of the mass $m$ moving on the half-line $-\infty<x<0$ and satisfying the Schr\"{o}dinger equation for the wave function $\Psi_n(x)$ and energy $E_n$, $n=0,1,2,\ldots$,
\begin{equation}\label{Schrodinger1}
-\frac{\hbar^2}{2m}\frac{d^2}{dx^2}\Psi_n(x)+V(x)\Psi_n(x)=E_n\Psi_n(x)
\end{equation}
for the finite nonzero extrapolation length $\Lambda$ is determined from the transcendental equation \cite{Olendski1}
\begin{equation}\label{EigenValueRobin1}
\left(\frac{2me\mathscr{E}}{\hbar^2}\right)^{1/3}{\rm Ai}'\!\left(-\left[\frac{2m}{(e\hbar\mathscr{E})^2}\right]^{1/3}\!\!E_n\right)-\frac{1}{\Lambda}\,{\rm Ai}\!\left(-\left[\frac{2m}{(e\hbar\mathscr{E})^2}\right]^{1/3}\!\!E_n\right)=0.
\end{equation}
Here, $\bf n$ is an inward unit normal to the interface $\cal S$, $V(x)$ is an electrostatic potential
\begin{equation}\label{Potential1}
V(x)=\left\{\begin{array}{cc}
-e\mathscr{E}x,&-\infty<x\leq0\\
\infty,&x>0
\end{array}\right.
\end{equation}
creating the field $\mathscr{E}>0$ that pushes the particle (for definiteness, we will talk about the electron with its negative charge $-e$) closer to the surface, ${\rm Ai}(x)$ is Airy function \cite{Abramowitz1,Vallee1} and prime denotes its derivative with respect to argument $x$. For $\Lambda=0$ ($\Lambda=\infty$) the BC from Eq.~\eqref{Robin1} simplifies to the Dirichlet (Neumann) requirement $\Psi(0)=0$ [$\Psi'(0)=0$] with the energy spectrum directly following from Eq.~\eqref{EigenValueRobin1}:
\begin{equation}\label{EigenValueDirNeumann1}
E_n^{\left\{_N^D\right\}}=-\left\{\begin{array}{cc}a_{n+1}\\a_{n+1}'\end{array}\right\}\left(\frac{e^2\hbar^2\mathscr{E}^2}{2m}\right)^{1/3},
\end{equation}
where the negative $a_n$ and $a_n'$ are $n$th roots, $n=1,2,\ldots$, of the equations ${\rm Ai}(x)=0$ or ${\rm Ai}'(x)=0$, respectively \cite{Abramowitz1,Vallee1}. For the small nonzero parameter $2me\mathscr{E}|\Lambda|^3/\hbar^2$ the Taylor expansion of Eq.~\eqref{EigenValueRobin1} shows that the first-order correction to the Dirichlet spectrum is equal to the voltage drop across the Robin distance:
\begin{subequations}\label{AsymptoticEnergies1}
\begin{align}
&E_n^{R\mp}=-\left\{\!\!\begin{array}{c}
a_n\\a_{n+1}
\end{array}\!\!\right\}\left(\frac{e^2\hbar^2\mathscr{E}^2}{2m}\right)^{1/3}+e\mathscr{E}\Lambda,\,\left\{\!\!\begin{array}{c}
n=1,2,3,\ldots\\n=0,1,2,\ldots
\end{array}\!\!\right\},\nonumber\\
\label{AsymptoticEnergies1_DirichletN}
&\frac{2me\mathscr{E}|\Lambda|^3}{\hbar^2}\ll1.\\
\intertext{The sign in the energy superscript corresponds to that of the extrapolation length. In addition, for the lowest level of the negative Robin wall one has:}
&E_0^{R-}=-\frac{\hbar^2}{2m|\Lambda|^2}\left[1-\frac{1}{2}\frac{2me\mathscr{E}|\Lambda|^3}{\hbar^2}+\frac{1}{8}\left(\frac{2me\mathscr{E}|\Lambda|^3}{\hbar^2}\right)^2\right],\nonumber\\
\label{AsymptoticEnergies1_Dirichlet0}
&\frac{2me\mathscr{E}|\Lambda|^3}{\hbar^2}\ll1.\\
\intertext{In the opposite limit the energies are:}
\label{AsymptoticEnergies1_Neumann}
&E_n^{R\mp}=-a_{n+1}'\left(\frac{e^2\hbar^2\mathscr{E}^2}{2m}\right)^{1/3}\left[1\mp\frac{1}{{a_{n+1}'}^{\!\!\!\!\!\!\!\!2}\,\,\,\,}\left(\frac{\hbar^2}{2me\mathscr{E}|\Lambda|^3}\right)^{1/3}\right],\nonumber\\
&n=0,1,2,\ldots,\quad\frac{\hbar^2}{2me\mathscr{E}|\Lambda|^3}\ll1.
\end{align}
\end{subequations}
Eq.~\eqref{AsymptoticEnergies1_Neumann} manifests that the energy admixture to the slightly distorted Neumann wall is of the order $1/\left(\Lambda\mathscr{E}^{1/3}\right)$. Hence, the varying electric field changes the energy spectrum in a very wide range. This is especially clearly seen after introducing the dimensionless scaling where all distances are measured in units of the magnitude of the Robin length $|\Lambda|$, energies -- in units of $\hbar^2/(2m|\Lambda|^2)$, and electric fields -- in units of $\hbar^2/(2me|\Lambda|^3)$. Then, the spectrum of the attractive Robin wall at the low voltages consists of the densely populated positive part
\begin{subequations}\label{AsymptoticEnergies2}
\begin{align}\label{AsymptoticEnergies2_1}
E_n^{R-}=-a_n\mathscr{E}^{2/3}-\mathscr{E},\quad n=1,2,\ldots,\,\mathscr{E}\ll1,
\intertext{with the energy  difference $\delta E_n^{R-}=E_{n+1}^{R-}-E_n^{R-}$ between the neighboring states decreasing for the growing $n$:}
\label{AsymptoticEnergies2_DeltaEnergies1}
\delta E_n^{R-}=\left(\frac{2}{3}\frac{\pi^2}{n}\mathscr{E}^2\right)^{1/3},\quad \mathscr{E}\ll1,\,n\gg1,
\intertext{and the negative-energy level that is separated from the quasi continuum by the almost unit gap \cite{Olendski1}:}
\label{AsymptoticEnergies2_Negative0}
E_0^{R-}=-1+\frac{1}{2}\,\mathscr{E}-\frac{1}{8}\,\mathscr{E}^2,\quad\mathscr{E}\ll1.
\end{align}
\end{subequations}
This highly nonuniform distribution of the levels on the energy axis leads, as shown below, to the spectacular thermodynamic properties, such as, for example, the gigantic growth of the heat capacity $c_V$, which are calculated on the basis of the knowledge of the spectrum $E_n$. Similar to the tilted quantum well (QW) with miscellaneous combinations of the Dirichlet and Neumann BCs at the opposite walls \cite{Olendski2,Olendski3}, the analysis below addresses both canonical, Sec.~\ref{Sec_Canonical}, and two types of the grand canonical ensembles, Sec.~\ref{Sec_GrandCanonical}. Previous research on the BC influence on the thermodynamics of the low-dimensional nanostructures concentrated on the calculations of the bosonic systems only and was limited to the flat QWs either with periodic BCs or at $\Lambda=0$ and/or $\Lambda=\infty$ relevant for the study of liquid helium in pore geometries and thin film superconductors \cite{Krueger1,Sonin1,Greenspoon1,Zasada1,Goble1,Goble2,Goble3,Grossmann1,Pathria1,Greenspoon2,Barber1,Hasan1}. In a recent contribution, the author has shown \cite{Olendski3} that for the canonical distribution the permutation of the Dirichlet and Neumann BCs of a QW at $\mathscr{E}=0$ has a significant impact on its thermal properties; for example, the heat capacity of the pure Neumann configuration exhibits on the temperature axis a conspicuous maximum accompanied by the adjacent minimum that are absent for any other BC geometry. Its physical explanation invoked an analysis of the corresponding structures of the associated spectra what mathematically reduced to correct handling of Theta functions. A remarkable feature of the fermionic QW is,  at any BC combination in zero fields, a  salient maximum of the specific heat for one particle and its absence for any other number of corpuscles. It was shown that the voltage applied to the QW leads to the increase of the heat capacity and formation of the new or modification of the existing extrema what is qualitatively described by the influence of the associated electric potential. The main message of the present research is to show on the example of one wall that the impact of the field for either type of the statistical ensemble has much more drastic consequences for the Robin BC, especially for that with $\Lambda<0$. Below, in addition to the dimensionless units introduced above, we will measure the heat capacity, if not specified otherwise, in units of the Boltzmann constant $k_B$, and particle kinetic momentum $p$ -- in units of $\hbar/|\Lambda|$.

\section{Canonical Ensemble}\label{Sec_Canonical}
The basic quantity in the description of the system that is in the thermal equilibrium with the much larger bath is the partition function defined as
\begin{equation}\label{PartitionFunction1}
Z(\mathscr{E};\beta)=\sum_ne^{-\beta E_n(\mathscr{E})}
\end{equation}
with the summation running over all possible quantum states, the parameter $\beta$ being (in regular, unnormalized units) $\beta=1/(k_BT)$, and $T$ is the thermodynamic temperature of the bath. For example, the probability $w_n$ of finding particle in the state $n$ is a function of the temperature and the energy $E_n(\mathscr{E})$, which, in turn, depends on the parameter $\mathscr{E}$, as discussed in the Introduction:
\begin{equation}\label{ProbabilityCanonical1}
w_n=\frac{1}{Z}\,e^{-\beta E_n},
\end{equation}
and the mean value $\langle{\cal I}\rangle_{can}$ of any physical quantity $\cal I$ is calculated as
\begin{equation}\label{CanonicalMeanValue1}
\langle{\cal I}\rangle_{can}=\frac{1}{Z}\sum_nw_n{\cal I}_n=\frac{\sum_{n}{\cal I}_ne^{-\beta E_n}}{\sum_{n}e^{-\beta E_n}}.
\end{equation}
The mean energy
\begin{subequations}\label{CanonicalMeanEnergy1}
\begin{align}\label{CanonicalMeanEnergy1_1}
\langle E\rangle_{can}(\mathscr{E};\beta)&=\frac{\sum_{n=0}^\infty E_ne^{-\beta E_n}}{\sum_{n=0}^\infty e^{-\beta E_n}}
\intertext{can equivalently be represented as}
\label{CanonicalMeanEnergy1_2}
\langle E\rangle_{can}&=-\frac{\partial}{\partial\beta}\ln Z
\end{align}
\end{subequations}
while the heat capacity $c_V$, i.e., a work that has to be done to change the temperature of the system by one degree
\begin{subequations}\label{HeatCapacity1}
\begin{align}\label{HeatCapacity1_1}
c_V=\frac{\partial}{\partial T}\langle E\rangle=-k_B\beta^2\frac{\partial}{\partial\beta}\langle E\rangle
\intertext{(above regular, unnormalized units have been used again), for the canonical ensemble is expressed with the help of the fluctuation-dissipation theorem \cite{Dalarsson1}:}
\label{HeatCapacity1_2}
c_{can}(\beta,\mathscr{E})=\beta^2\left(\langle E^2\rangle_{can}-\langle E\rangle_{can}^2\right).
\end{align}
\end{subequations}
For $N$ noninteracting corpuscles in the system, the right-hand sides of Eqs.~\eqref{CanonicalMeanValue1} -- \eqref{HeatCapacity1} have to be multiplied by $N$.

\begin{figure}
\centering
\includegraphics[width=\columnwidth]{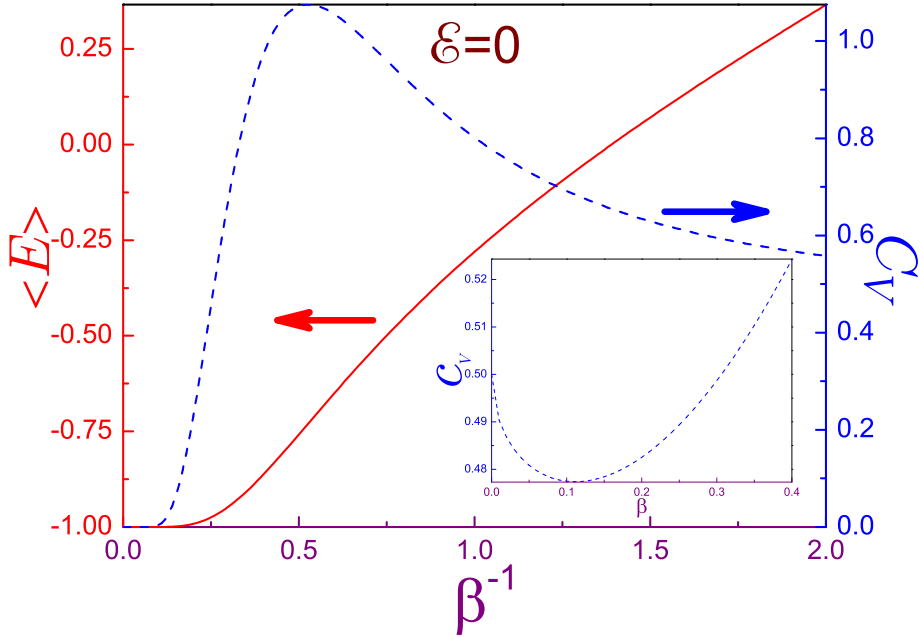}
\caption{\label{HeatCapacityCanonicalNoFieldsFig1}
Zero-field canonical mean energy $\langle E\rangle$ (solid line, left axis) and heat capacity $c_V$ (dashed curve, right axis) of the attractive Robin wall as a function of temperature $\beta^{-1}$. Inset shows the peculiarity of the heat capacity at the large temperatures.}
\end{figure}

Before discussing the electric field influence on the thermodynamics of the system, it does make sense to analyze first a zero-voltage case. For the nonnegative extrapolation lengths characterized by the positive-energy continuous spectrum only the summation in the partition function, Eq.~\eqref{PartitionFunction1}, is replaced by the integration over the momentum $p$:
\begin{equation}\label{PartitionFunctionDNRplus}
Z^{D,N,R+}(0;\beta)=\int_{-\infty}^\infty e^{-\beta p^2}dp=\left(\frac{2\pi}{\beta}\right)^{1/2},
\end{equation}
what leads to the same expressions of the mean energy and temperature independent specific heat as for the free space \cite{Dalarsson1}:
\begin{eqnarray}\label{CanonicalMeanEnergyZeroFieldDNRplus}
\langle E\rangle_{can}^{D,N,R+}(0;\beta)&=&\frac{1}{2\beta}\\
\label{HeatCapacityZeroFieldDNRplus}
c_{can}^{D,N,R+}(0;\beta)&=&\frac{1}{2}.
\end{eqnarray}
Situation is different for the attractive Robin wall. The existence of the bound state modifies the partition function as
\begin{equation}\label{PartitionFunctionRminus}
Z^{R-}(0;\beta)=e^\beta+\left(\frac{2\pi}{\beta}\right)^{1/2},
\end{equation}
what results in the following expression for the mean energy:
\begin{subequations}\label{CanonicalMeanEnergyZeroFieldRminus1}
\begin{align}\label{CanonicalMeanEnergyZeroFieldRminus1_1}
\langle E\rangle_{can}^{R-}(0;\beta)=\frac{-e^\beta+\frac{1}{2}\left(\frac{2\pi}{\beta^3}\right)^{1/2}}{e^\beta+\left(\frac{2\pi}{\beta}\right)^{1/2}}.
\intertext{Its asymptotic cases for the low}
\langle E\rangle_{can}^{R-}(0;\beta)=-1+\left(\frac{2}{\pi\beta}\right)^{1/2}\left(1+\frac{1}{2\beta}\right)e^{-\beta}\nonumber\\
\label{CanonicalMeanEnergyZeroFieldRminus1_3}
-\frac{2\pi}{\beta}\left(1+\frac{1}{2\beta}\right)e^{-2\beta}+\ldots,\quad\beta\rightarrow\infty,
\intertext{and high temperatures}
\langle E\rangle_{can}^{R-}(0;\beta)=\frac{1}{2\beta}-\frac{1}{4}\left(\frac{2}{\pi\beta}\right)^{1/2}+\frac{1}{4\pi}\nonumber\\
\label{CanonicalMeanEnergyZeroFieldRminus1_2}
-\frac{1}{2}\left(\frac{2}{\pi}\beta\right)^{1/2}\left(\frac{3}{2}+\frac{1}{4\pi}\right)+\ldots,\quad\beta\rightarrow0,
\end{align}
\end{subequations}
allow with the help of Eq.~\eqref{HeatCapacity1_1} to calculate the specific heat $c_V$ in the same limits as:
\begin{subequations}\label{HeatCapacityZeroFieldDNRminus1}
\begin{eqnarray}
c_{can}^{R-}(0;\beta)&=&\left(\frac{2}{\pi\beta}\right)^{1/2}\left(\frac{3}{4}+\beta+\beta^2\right)e^{-\beta}\nonumber\\
\label{HeatCapacityZeroFieldDNRminus1_2}
&-&\frac{2}{\pi\beta}\left(1+2\beta+2\beta^2\right)e^{-2\beta}+\ldots,\quad\beta\rightarrow\infty\\
c_{can}^{R-}(0;\beta)&=&\frac{1}{2}-\frac{1}{8}\left(\frac{2}{\pi}\right)^{1/2}\beta^{1/2}\nonumber\\
\label{HeatCapacityZeroFieldDNRminus1_1}
&+&\frac{1}{4}\left(\frac{2}{\pi}\right)^{1/2}\left(\frac{3}{2}+\frac{1}{4\pi}\right)\beta^{3/2}+\ldots,\quad\beta\rightarrow0.
\end{eqnarray}
\end{subequations}

The variation of the mean energy and specific heat with the temperature $\beta^{-1}$ is shown in Fig.~\ref{HeatCapacityCanonicalNoFieldsFig1}. It is seen that the initial warming of the extremely cold wall has an exponentially negligible effect on both parameters, in accordance with Eqs.~\eqref{CanonicalMeanEnergyZeroFieldRminus1_3} and \eqref{HeatCapacityZeroFieldDNRminus1_2}. This is explained by the smallness of the thermal energy $1/(2\beta)$ with respect to the gap between the bound state and the positive energy continuum. Subsequent growth of the temperature causes a monotonic increase of the mean energy that at $\beta\rightarrow0$ approaches asymptotically the value $1/(2\beta)$ of the free particle, as Eq.~\eqref{CanonicalMeanEnergyZeroFieldRminus1_2} also demonstrates. The avalanche rise of the heat capacity starts at the smaller temperatures as compared to that of the mean energy. The specific heat after the steep surge reaches at $\beta_{max}^{-1}=0.5260$ ($\beta_{max}=1.9010$) the maximum of $c_{max}=1.0752$. Its following decrease terminates at $\beta_{min}^{-1}=8.9684$ ($\beta_{min}=0.1115$) where the minimum $c_{min}=0.4774$ is located. The higher temperatures lead to the smooth nearing of the heat capacity to the free-electron configuration of $1/2$. Observe that these extrema and their positions can be with the decent accuracy extracted from the asymptotic expansions from Eqs.~\eqref{HeatCapacityZeroFieldDNRminus1}; for example, a zeroing of  the $\beta$-derivative of Eq.~\eqref{HeatCapacityZeroFieldDNRminus1_1} yields the approximate location of the minimum $c_{min}^{appr}=0.4784$ at
$$\beta_{min}^{appr}=\frac{1}{9+\frac{3}{2\pi}}=0.1055,$$ which coincide quite well with the values provided above.

\begin{figure}
\centering
\includegraphics[width=\columnwidth]{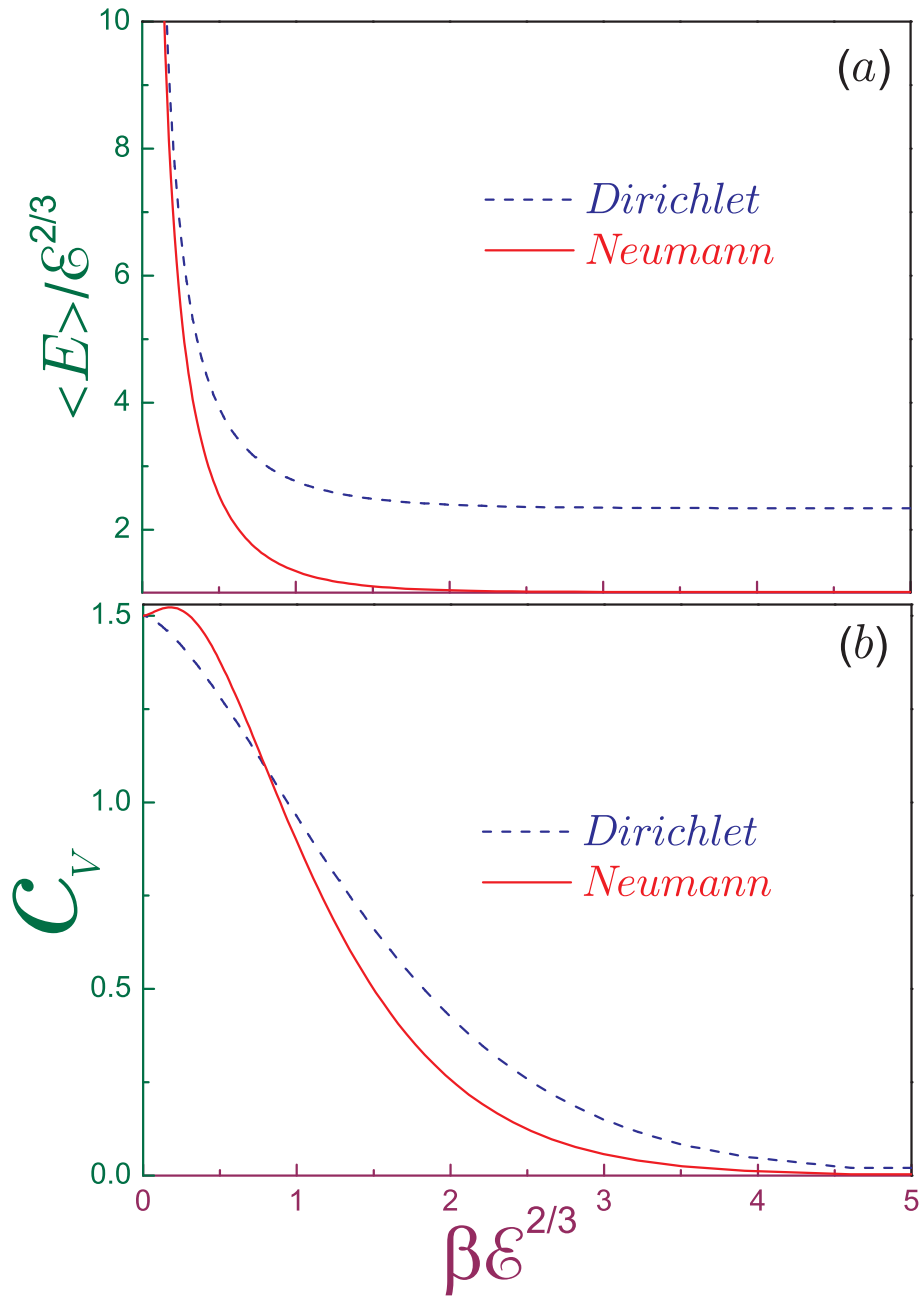}
\caption{\label{CanonicalEnergyHeatCapDirNeuFig1}
Universal curves of canonical (a) ratio $\left.\left\langle E^{_N^D}\right\rangle\right/\mathscr{E}^{2/3}$ and (b) heat capacity $c_V$ versus the parameter $\beta\mathscr{E}^{2/3}$ for the Dirichlet (dashed lines) and Neumann (solid lines) walls.}
\end{figure}

Thus, the formation by the attractive Robin wall of the split-off negative-energy bound state strongly modifies its thermodynamic properties in comparison with the interfaces with the nonnegative $\Lambda$; for example, it leads to the nonmonotonic dependence of the heat capacity on the temperature, as just shown above. The difference is greatly enhanced upon application of the electric field. First, let us note that the classical "potential" partition function \cite{Dalarsson1}
\begin{equation}\label{PotentialPartitionFunction1}
Z_{pot}=\int_{-\infty}^\infty e^{-\beta V(x)}dx
\end{equation}
for the configuration from Eq.~\eqref{Potential1} becomes:
\begin{equation}\label{PotentialPartitionFunction2}
Z_{pot}=\frac{1}{\beta\mathscr{E}},
\end{equation}
while a thermally averaged value of the potential energy
\begin{equation}
\langle V(x)\rangle=\frac{1}{Z_{pot}}\int_{-\infty}^\infty V(x)e^{-\beta V(x)}dx
\end{equation}
for the Robin wall in the electric field is just the heat quantum:
\begin{equation}\label{PotentialAveraged1}
\langle V(x)\rangle=\frac{1}{\beta}.
\end{equation}
Then, the potential contribution to the heat capacity
\begin{equation}\label{HeatCapacityPotentialContribution1}
c_{pot}=-\beta^2\frac{\partial}{\partial\beta}\langle V(x)\rangle
\end{equation}
turns to unity what means that the total specific heat at the high temperatures when the classical expressions from Eqs.~\eqref{PotentialPartitionFunction1}--\eqref{HeatCapacityPotentialContribution1} are only applicable to the quantum case is equal to $3/2$.

Expressions for the Dirichlet and Neumann mean energies
\begin{equation}\label{CanonicalMeanEnergyDN1}
\left\langle E^{_N^D}\right\rangle=\frac{\mathscr{E}^{2/3}}{Z^{_N^D}}\sum_{n=1}^\infty b_n^{_N^D}\exp\!\left(-b_n^{_N^D}\beta\mathscr{E}^{2/3}\right)
\end{equation}
with the associated partition functions
\begin{equation}\label{PartitionFunctionDN1}
Z^{_N^D}=\sum_{n=1}^\infty \exp\left(-b_n^{_N^D}\beta\mathscr{E}^{2/3}\right)
\end{equation}
and coefficients
\begin{equation}\label{CoefficientsBn}
b_n^{_N^D}=\left\{\begin{array}{c}
-a_n\\-a_n'
\end{array}\right.
\end{equation}
show that the corresponding heat capacities $c_V^{_N^D}$ and the values $\left.\left\langle E^{_N^D}\right\rangle\right/\mathscr{E}^{2/3}$ are functions of one variable $y\equiv\beta\mathscr{E}^{2/3}$ only. In particular, for the large values of this parameter, $y\gg1$, one has:
\begin{subequations}\label{AsymptoteCanonicalDNlargeY1}
\begin{eqnarray}
\frac{\langle E\rangle}{\mathscr{E}^{2/3}}&=&b_1+(b_2-b_1)e^{-(b_2-b_1)y}+(b_3-b_1)e^{-(b_3-b_1)y}\nonumber\\
\label{AsymptoteCanonicalDNlargeY1_Energy}
&+&(b_2-b_1)^2e^{-2(b_2-b_1)y}+\ldots\\
c_V&=&y^2\left[(b_2-b_1)^2e^{-(b_2-b_1)y}+(b_3-b_1)^2e^{-(b_3-b_1)y}\right.\nonumber\\
\label{AsymptoteCanonicalDNlargeY1_Capacity}
&+&\left.2(b_2-b_1)^3e^{-2(b_2-b_1)y}+\ldots\right].
\end{eqnarray}
\end{subequations}
To find these functions performance in the opposite limit of $y\ll1$, one employs the behavior of the Airy zeros coefficients at the large index \cite{Abramowitz1,Vallee1}
\begin{equation}\label{CoefficientsA1}
\left\{\begin{array}{c}
a_n\\a_n'
\end{array}\right\}=-\left[\frac{3\pi}{8}\left(4n-\left\{\begin{array}{c}
1\\3
\end{array}\right\}\right)\right]^{2/3},\quad n\gg1,
\end{equation}
which produce a reasonably good accuracy even for $n=1$ \cite{Olendski4}, and the asymptotic expansion of the infinite series
\begin{subequations}\label{AsymptoticSeries1}
\begin{align}
\sum_{n=0}^\infty e^{-t(n+d)^\alpha}=\frac{\Gamma(1+1/\alpha)}{t^{1/\alpha}}-\left(d-\frac{1}{2}\right)\nonumber\\
\label{AsymptoticSeries1_1}
+\frac{\left|d-1/2\right|^{1+1/\alpha}}{1+\alpha}t-\ldots,\quad t\rightarrow0
\intertext{and its derivative with respect to the variable $t$:}
\sum_{n=0}^\infty(n+d)^\alpha e^{-t(n+d)^\alpha}=\frac{\Gamma(1+1/\alpha)}{\alpha t^{1+1/\alpha}}-\frac{\left|d-1/2\right|^{1+1/\alpha}}{1+\alpha}\nonumber\\
\label{AsymptoticSeries1_2}
+\frac{\left|d-1/2\right|^{2+1/\alpha}}{1+2\alpha}t-\ldots,\quad t\rightarrow0,
\end{align}
\end{subequations}
with $\Gamma(\alpha)$ being $\Gamma$-function \cite{Abramowitz1}. As a result, one obtains the following expressions for the heat capacities at $y\ll1$:
\begin{equation}\label{AsymptoteCanonicalDNsmallY1}
c^{{}_N^D}=\frac{3}{2}\left(1\mp\frac{\pi^{1/2}}{4}\beta^{3/2}\mathscr{E}\right),\quad\beta\mathscr{E}^{2/3}\ll1.
\end{equation}
Observe that the Neumann specific heat approaches from above the infinite temperature limit of $3/2$ what means that it has a maximum on the $y$ axis. This qualitative conclusion is confirmed by the exact calculations shown in Fig.~\ref{CanonicalEnergyHeatCapDirNeuFig1}, which exhibits the universal dependencies of the heat capacity and the ratio $\left.\left\langle E^{_N^D}\right\rangle\right/\mathscr{E}^{2/3}$ on the variable $y$ for the Dirichlet and Neumann surfaces. The specific heat $c_V^N$ reaches a shallow maximum of $c_{max}^N=1.522$ at $y_{max}=0.175$. It is seen that at the large $y$ the Neumann heat capacity fades faster since the corresponding difference $a_1'-a_2'=2.2294$ in Eq.~\eqref{AsymptoteCanonicalDNlargeY1_Capacity} is greater than its Dirichlet counterpart $a_1-a_2=1.7498$ \cite{Abramowitz1,Vallee1}. The Dirichlet mean energy lies above the Neumann one since, for example, in the same limit of $y\gg1$ they are mainly determined, as Eq.~\eqref{AsymptoteCanonicalDNlargeY1_Energy} demonstrates, by the first Airy zeros $a_1=-2.3381$ and  $a_1'=-1.0188$, respectively.

\begin{figure}
\centering
\includegraphics[width=\columnwidth]{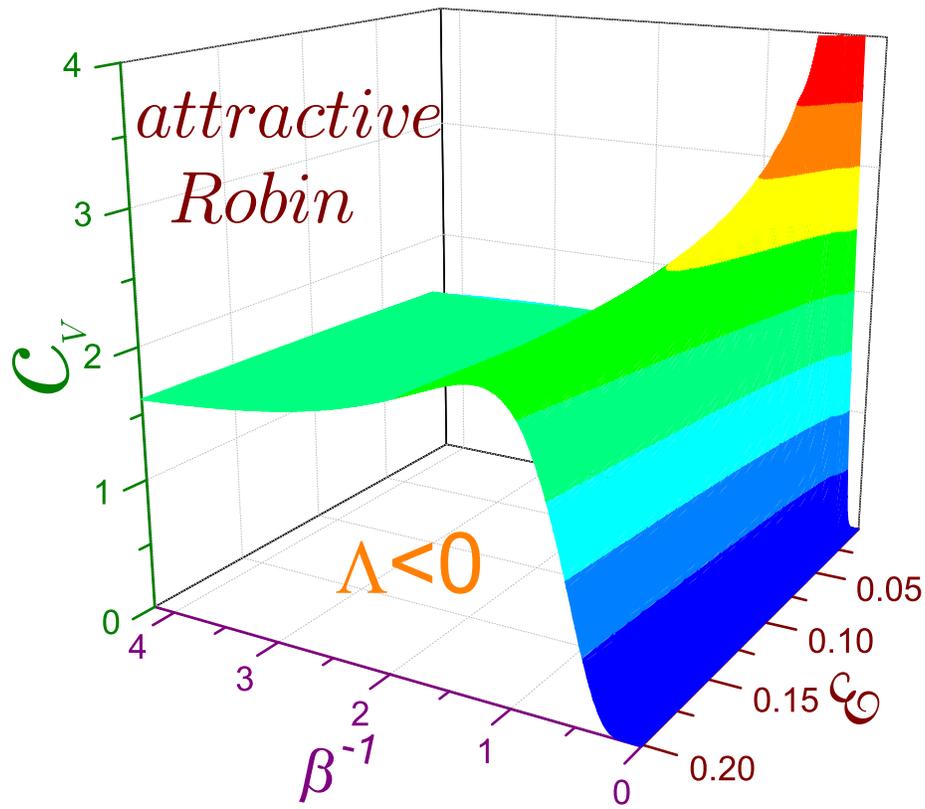}
\caption{\label{HeatCapacityCanonicalFig1}
Canonical specific heat $c_V^{R-}$ of the attractive Robin wall in terms of the temperature $\beta^{-1}$ and electric field $\mathscr{E}$.}
\end{figure}

\begin{figure}
\centering
\includegraphics[width=\columnwidth]{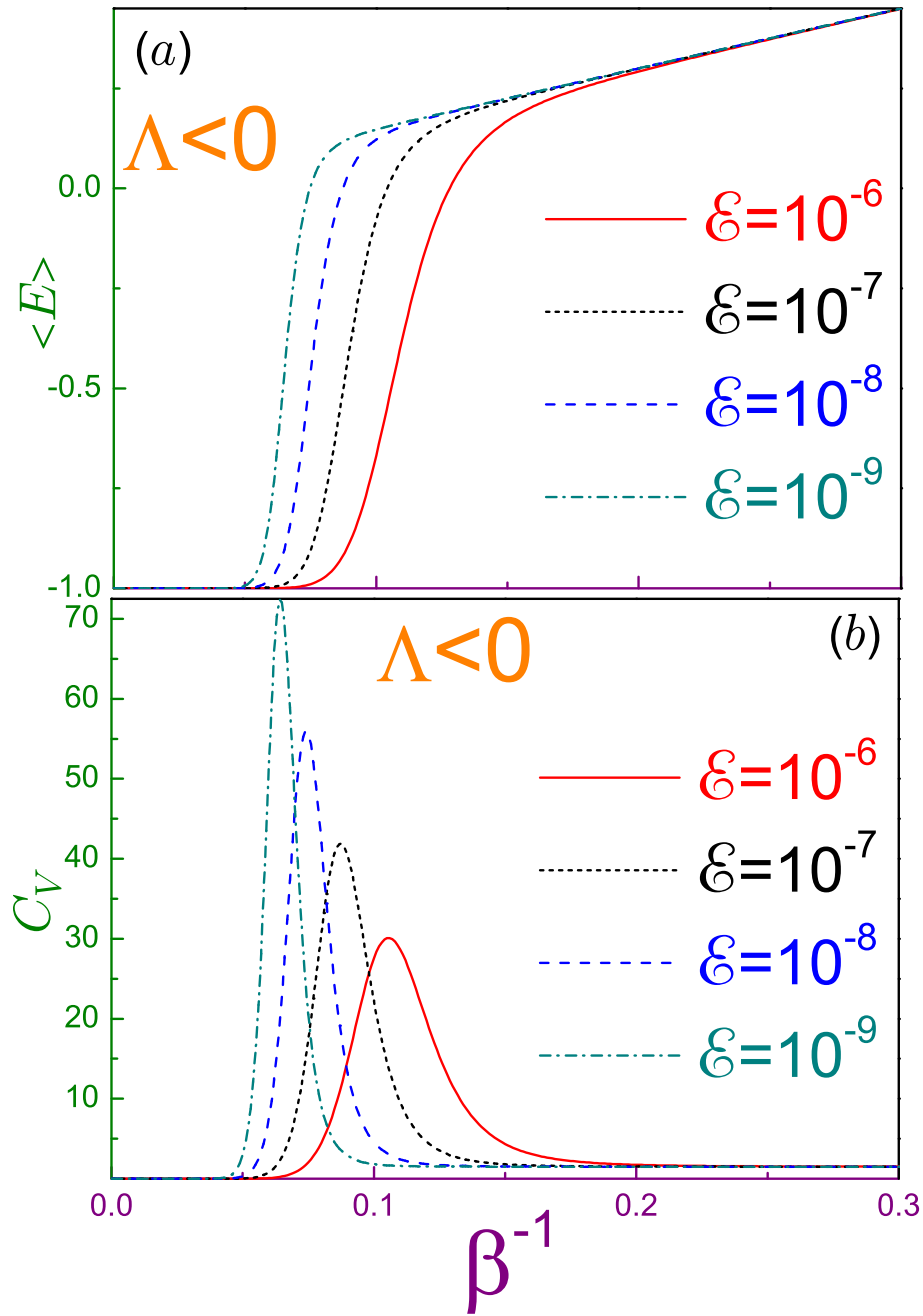}
\caption{\label{EnergyHeatcapacityNegativeLambdaCanonicalSmallFieldsFig1}
Canonical (a) mean energy $\langle E^{R-}\rangle$ and (b) specific heat $c_V^{R-}$ of the attractive Robin wall as functions of the low temperature $\beta^{-1}$ at several extremely weak electric fields where solid lines are for $\mathscr{E}=10^{-6}$, dotted curves -- for $\mathscr{E}=10^{-7}$, dashed lines -- for $\mathscr{E}=10^{-8}$, and dash-dotted curves are for $\mathscr{E}=10^{-9}$.}
\end{figure}

The most dramatic feature of the heat capacity of the negative Robin wall shown in Fig.~\ref{HeatCapacityCanonicalFig1} is its colossal increase at the weak fields and low voltage-dependent temperatures. To explain this phenomenon from a mathematical point of view, it suffices to keep in Eqs.~\eqref{AsymptoticEnergies2_1} and \eqref{AsymptoticEnergies2_Negative0} of the corresponding energy spectrum the leading terms only. Next, assuming that, in addition to the condition $\mathscr{E}\ll1$, the requirement $\beta\mathscr{E}^{2/3}\ll1$ also holds (its validity for our range of interest will be justified below), one invokes the expression for the coefficients $a_n$ from Eq.~\eqref{CoefficientsA1} and the expansions from Eqs.~\eqref{AsymptoticSeries1}. As a result, the mean energy in these limits reads:
\begin{equation}\label{MeanEnergyLimit1}
\left\langle E^{R-}\right\rangle=\frac{-e^\beta+\frac{3}{4\pi^{1/2}}\frac{1}{\mathscr{E}\beta^{5/2}}}{e^\beta+\frac{1}{2\pi^{1/2}}\frac{1}{\mathscr{E}\beta^{3/2}}},\quad\mathscr{E},\,\beta\mathscr{E}^{2/3}\rightarrow0.
\end{equation}
First, for the large temperatures it simplifies to
\begin{subequations}
\begin{align}\label{MeanEnergyAsympt1}
\left\langle E^{R-}\right\rangle=\frac{3}{2}\frac{1}{\beta}-3\pi^{1/2}\mathscr{E}\beta^{1/2}-5\pi^{1/2}\mathscr{E}\beta^{3/2}+\ldots,\quad\beta,\mathscr{E}\rightarrow0,
\intertext{what results in the heat capacity:}
\label{HeatCapacityAsymptotics2}
c_V^{R-}=\frac{3}{2}+\frac{3}{2}\pi^{1/2}\mathscr{E}\beta^{3/2}+\frac{15}{2}\pi^{1/2}\mathscr{E}\beta^{5/2}+\ldots,\quad\beta,\mathscr{E}\rightarrow0.
\end{align}
\end{subequations}
Note that the last expression remains intact if the terms linear in the field are retained in Eqs.~\eqref{AsymptoticEnergies2_1} and \eqref{AsymptoticEnergies2_Negative0}. This voltage-dependent nearing from above to the high-temperature limit is clearly seen in Fig.~\ref{HeatCapacityCanonicalFig1}. Next, the mean energy turns to zero at the temperature $\beta_{\langle E\rangle=0}$, which is found from equation:
\begin{equation}\label{TemperatureOfZeroEnergy1}
\beta^{5/2}e^\beta=\frac{3}{4\pi^{1/2}}\frac{1}{\mathscr{E}}.
\end{equation}
Its solution is expressed via the Lambert $W$ function \cite{Corless1,Veberic1,Houari1}
\begin{equation}\label{TemperatureOfZeroEnergy2}
\beta_{\langle E\rangle=0}=\frac{5}{2}\,W\!\!\left(\frac{9^{1/5}16^{4/5}}{\pi^{1/5}\mathscr{E}^{2/5}}\right).
\end{equation}
Asymptotics of the Lambert function \cite{Corless1}
\begin{equation}\label{LambertAsymptotics1}
W(x)\rightarrow\ln\frac{x}{\ln x}+\ldots,\quad x\rightarrow\infty,
\end{equation}
shows that at the very weak (but nonzero) fields the temperature at which the zero mean energy is achieved, is inversely proportional to the logarithm of the applied voltage:
\begin{equation}\label{TemperatureOfZeroEnergy3}
\beta_{\langle E\rangle=0}\simeq-\ln\mathscr{E},\quad\mathscr{E}\rightarrow0.
\end{equation}
Eq.~\eqref{TemperatureOfZeroEnergy3} manifests that the mean energy for the decreasing $\mathscr{E}$ becomes a steeper function of the low temperature. This is exemplified in panel (a) of Fig.~\ref{EnergyHeatcapacityNegativeLambdaCanonicalSmallFieldsFig1} that shows the thermally averaged energies and heat capacities for several ultra weak electric intensities as functions of the temperature. Rapid growth of $\langle E^{R-}\rangle$ means, according to Eq.~\eqref{HeatCapacity1_1}, larger values of the heat capacity whose expression in the same limit reads 
\begin{equation}\label{HeatCapacityAsymptotics1}
c_V^{R-}=\frac{1}{2}\,\frac{3+\pi^{1/2}\mathscr{E}\beta^{3/2}e^\beta(4\beta^2+12\beta+15)}{\left(1+2\pi^{1/2}\mathscr{E}\beta^{3/2}e^\beta\right)^2},\quad\beta\mathscr{E}^{2/3}\rightarrow0.
\end{equation}
Its asymptotics for the large temperatures, which approaches, as expected, the value of $3/2$, was derived above, Eq.~\eqref{HeatCapacityAsymptotics2}. Taking a derivative $\partial_\beta$ of Eq.~\eqref{HeatCapacityAsymptotics1}, equating it to zero and keeping only the largest power of $\beta$ leads to the following expression for the temperature $\beta_{max}$ of the maximum of the specific heat:
\begin{equation}\label{TemperatureOfMaximum1}
\beta_{max}=\frac{3}{2}\,W\!\!\left(\frac{4^{2/3}}{6\pi^{1/3}\mathscr{E}^{2/3}}\right).
\end{equation}
Invoking the asymptotics from Eq.~\eqref{LambertAsymptotics1} produces:
\begin{equation}\label{TemperatureOfMaximum2}
\beta_{max}=\frac{3}{2}\ln\frac{4^{2/3}}{6\pi^{1/3}\mathscr{E}^{2/3}\ln\frac{4^{2/3}}{6\pi^{1/3}\mathscr{E}^{2/3}}},
\end{equation}
what essentially is the same dependence (with the different proportionality coefficient) as that from Eq.~\eqref{TemperatureOfZeroEnergy3}. Note that in this way our starting assumption of $\beta\mathscr{E}^{2/3}\rightarrow0$ is automatically satisfied. Corresponding peak of the heat capacity $c_{max}$ is calculated as
\begin{equation}\label{MaximumHeatCapacity1}
c_{max}^{R-}=\frac{1}{4}\beta_{max}^2.
\end{equation}
Eq.~\eqref{MaximumHeatCapacity1} and \eqref{TemperatureOfMaximum2} show that the maximum of the specific heat is proportional basically to $\ln^2\mathscr{E}$. This growth of $c_{max}^{R-}$ and its shift to the lower temperatures are seen in Figs.~\ref{HeatCapacityCanonicalFig1} and \ref{EnergyHeatcapacityNegativeLambdaCanonicalSmallFieldsFig1}(b). They also prove that the half width of this resonance narrows at the smaller electric intensities. 

\begin{figure}
\centering
\includegraphics[width=\columnwidth]{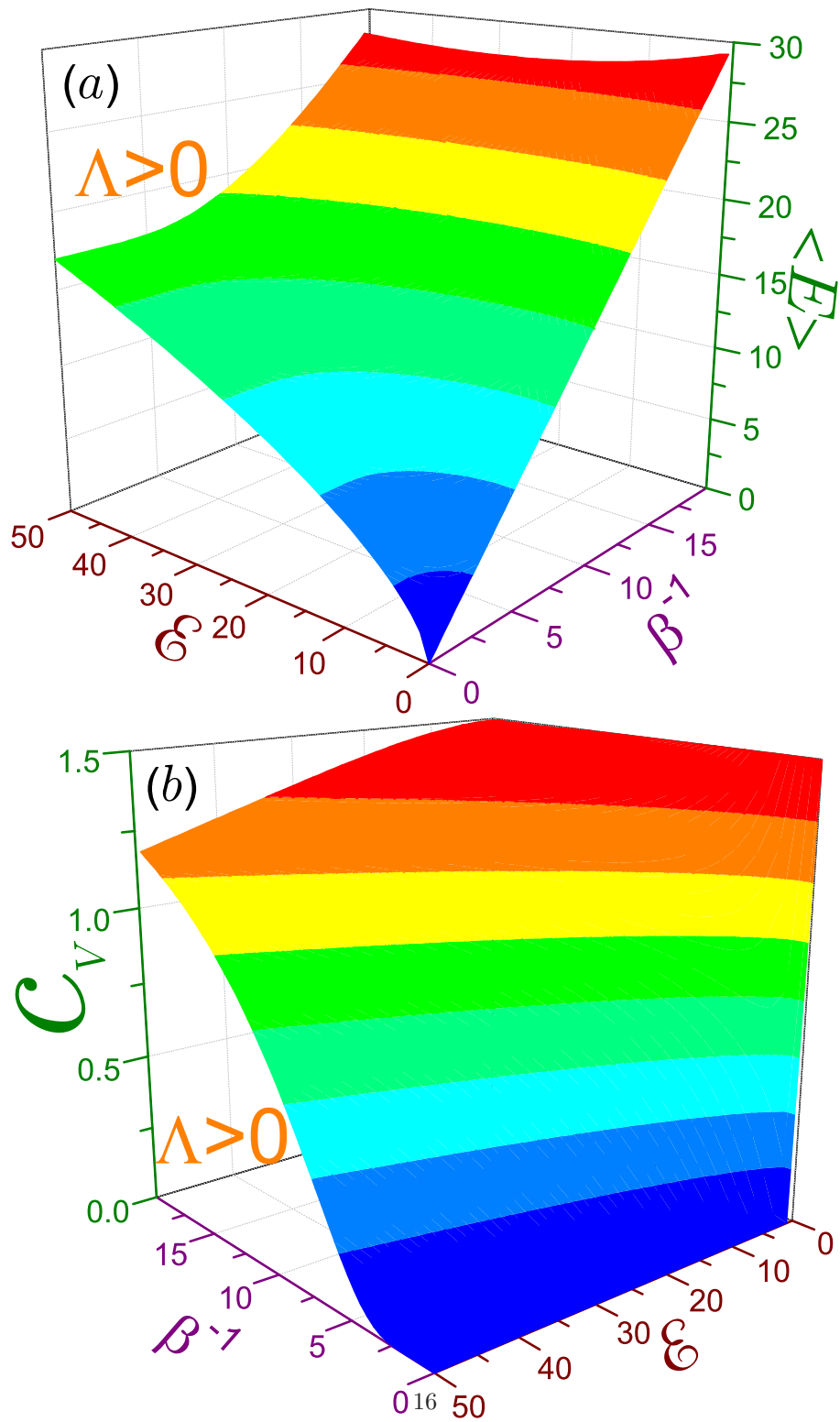}
\caption{\label{HeatCapacityCanonicalPositiveLambdaFig1}
Canonical (a) mean energy $\langle E^{R+}\rangle$ and (b) specific heat $c_V^{R+}$ of the repulsive Robin wall as functions of the temperature $\beta^{-1}$ and  electric field $\mathscr{E}$.}
\end{figure}

A physical reason of this giant enhancement of the heat capacity at the weak fields and extremely low voltage-dependent temperature stems from the peculiarities of the energy spectrum and its modification by the applied electric force. As discussed in the Introduction, at the weak intensities $\mathscr{E}$ the lowest lying level is split off from the quasi continuum of the positive-energy states. The difference 
\begin{equation}\label{EnergyDifference1}
\Delta_n^{R-}(\mathscr{E})=E_n^{R-}(\mathscr{E})-E_0^{R-}(\mathscr{E}),\quad n=1,2,\ldots,
\end{equation}
between the first excited, $n=1$, and ground state is in this regime practically the same as for many other higher lying levels with the ratio 
\begin{equation}\label{EnergyDifferenceRatio1}
{\cal R}_n^{R-}(\mathscr{E})=\frac{\Delta_n^{R-}(\mathscr{E})}{\Delta_1^{R-}(\mathscr{E})},\quad n=2,3,\ldots,
\end{equation}
turning to unity for the vanishing fields:
\begin{equation}\label{EnergyDifferenceRatio2}
{\cal R}_n^{R-}(\mathscr{E})=1+(a_1-a_n)\,\mathscr{E}^{2/3},\quad\mathscr{E}\ll1.
\end{equation}
Accordingly, the very low temperatures can promote the electron from its ground position not only to the first excited level but with about the same probability to a huge number of its counterparts what results in a steep increase of the heat capacity with its slope being more vertical for the weaker field. Further tiny warming of the structure can not sustain this very fast growth since the latter has to involve the levels with such enormously large indices $n$ for which the ratio ${\cal R}_n^{R-}$ considerably deviates from the unity what makes the transitions to them more difficult at this temperature. Consequently, the specific heat drops forming a characteristic sharp resonance with its peak increasing for the decreasing $\mathscr{E}$. The higher applied voltage splits wider the positive energy levels with the corresponding change of the ratio ${\cal R}$; for example, in the opposite limit of the large electric intensities it becomes a field independent quantity that noticeably grows with the quantum number $n$:
\begin{equation}\label{EnergyDifferenceRatio3}
{\cal R}_n^{R\pm}(\mathscr{E})=\frac{a_1'-a_{n+1}'}{a_1'-a_2'},\quad\mathscr{E}\gg1,
\end{equation}
as it follows from Eq.~\eqref{AsymptoticEnergies1_Neumann}. For that reason, in this case the low temperatures propel the electron mainly to the first excited state with the transitions to the higher levels being strongly suppressed due to their large energy separation from the one with $n=1$. They gradually can be populated one by one at the increasing temperatures. This change of the energy spectrum by the electric field smooths out the heat capacity and eventually at the growing $\mathscr{E}$ the resonance is transformed into the shallow Neumann maximum discussed above. As Fig.~\ref{HeatCapacityCanonicalFig1} demonstrates, the peak is strongly subdued already at $\mathscr{E}\sim0.2$. To underline the crucial role of the zero-field negative-energy bound state in the formation of the specific heat resonance, we provide in Fig.~\ref{HeatCapacityCanonicalPositiveLambdaFig1} the mean energy and the heat capacity of the repulsive Robin wall, which at any field is characterized by the positive spectrum only; for example, at the weak electric intensity the corresponding ratio ${\cal R}_n^{R+}$
\begin{equation}\label{EnergyDifferenceRatio4}
{\cal R}_n^{R+}=\frac{a_1-a_{n+1}}{a_1-a_2},\quad\mathscr{E}\ll1,
\end{equation}
similar to the high-field limit from Eq.~\eqref{EnergyDifferenceRatio3}, is a salient function of the index $n$. As a result, the heat capacity in the whole range of the electric forces is a monotonic function of the temperature, as panel (b) of Fig.~\ref{HeatCapacityCanonicalPositiveLambdaFig1} demonstrates. For the weaker voltages the high-temperature limit $3/2$ of the specific heat is achieved at the lower temperatures since the lesser energy differences
\begin{equation}\label{EnergyDifference2}
\Delta_n^{R+}(\mathscr{E})=(a_1-a_{n+1})\,\mathscr{E}^{2/3},\quad\mathscr{E}\ll1,\,n=1,2,\ldots,
\end{equation}
permit to occupy all the levels at the smaller $\beta^{-1}$. However, since the ratio from Eq.~\eqref{EnergyDifferenceRatio4} is strongly $n$ dependent, the heat capacity remains at these fields a monotonic function of the temperature too.

\section{Grand Canonical Ensemble}\label{Sec_GrandCanonical}
The quantum system that, in addition of being in the thermal balance with the external reservoir, can exchange the particles as well as energy with it, is said to be in the chemical equilibrium with the heat bath. Since the number $N$ of the constituent corpuscles varies, the properties of the structure strongly depend on it; for example, the chemical potential $\mu$ that is defined as the work that has to be done to change the number of the particles by one
\begin{equation}\label{ChemicalPotential1}
\mu=\left(\frac{\partial\langle E\rangle}{\partial N}\right)_{T,V},
\end{equation}
is determined from
\begin{equation}\label{NumberN_1}
N=\sum_{n=0}^\infty\frac{1}{e^{(E_n-\mu)\beta}\pm1},
\end{equation}
where the upper sign corresponds to the Fermi-Dirac (FD) distribution while the lower one describes Bose-Einstein (BE) systems. Knowledge of the mean energy
\begin{equation}\label{GrandCanonicalMeanEnergy1}
\langle E\rangle_{gc}=\sum_n\frac{E_n}{e^{(E_n-\mu)\beta}\pm1}
\end{equation}
allows one with the help of Eq.~\eqref{HeatCapacity1_1} to calculate the associated heat capacity $c_{gc}$ as
\begin{equation}\label{GrandCanonicalHeatCapacity1}
c_{gc}=\beta^2\sum_n\frac{E_n\left(E_n-\mu-\beta\frac{\partial\mu}{\partial\beta}\right)}{\left[e^{(E_n-\mu)\beta}\pm1\right]^2}\,e^{(E_n-\mu)\beta},
\end{equation}
where the chemical potential, which for the former ensemble is also frequently called the Fermi energy, is calculated from Eq.~\eqref{NumberN_1}, as just stated above, and its partial derivative with respect to the inverse temperature is found from the same expression with the help of the rule of differentiating of the implicit functions and is given as \cite{Olendski3}
\begin{equation}\label{Implicit1}
\beta\frac{\partial\mu}{\partial\beta}=\frac{\sum_n\frac{E_n-\mu}{\left[e^{(E_n-\mu)\beta}\pm1\right]^2}\,e^{(E_n-\mu)\beta}}{\sum_n\frac{1}{\left[e^{(E_n-\mu)\beta}\pm1\right]^2}\,e^{(E_n-\mu)\beta} }.
\end{equation}
Mathematical difference of the opposite signs in Eq.~\eqref{NumberN_1} has its physical consequence in the fact that each quantum level is occupied by at most one fermion while the arbitrary number of bosons can coexist in the same state. The FD distribution is applied for the system of the particles with the half-integer spin such as, for example, the electron in the free space or in the crystal lattice, while the boson statistics is used for the constituents with the integer spin, e.g., photons or Cooper pairs in superconductors. Thus, our remark from the Introduction that we consider the electron does not apply for the BE consideration where we will assume the abstract negatively charged particle. Also, for the half-integer spin fermions, say, the electron with its spin of $1/2$, the orbital states with the same energy can be occupied by the particles with the opposite projections of the spin $\pm1/2$. However, this fact is neglected in the discussion below where it is assumed that the number of the fermions for each energy $E_n$ is not larger than one.

\begin{figure}
\centering
\includegraphics[width=\columnwidth]{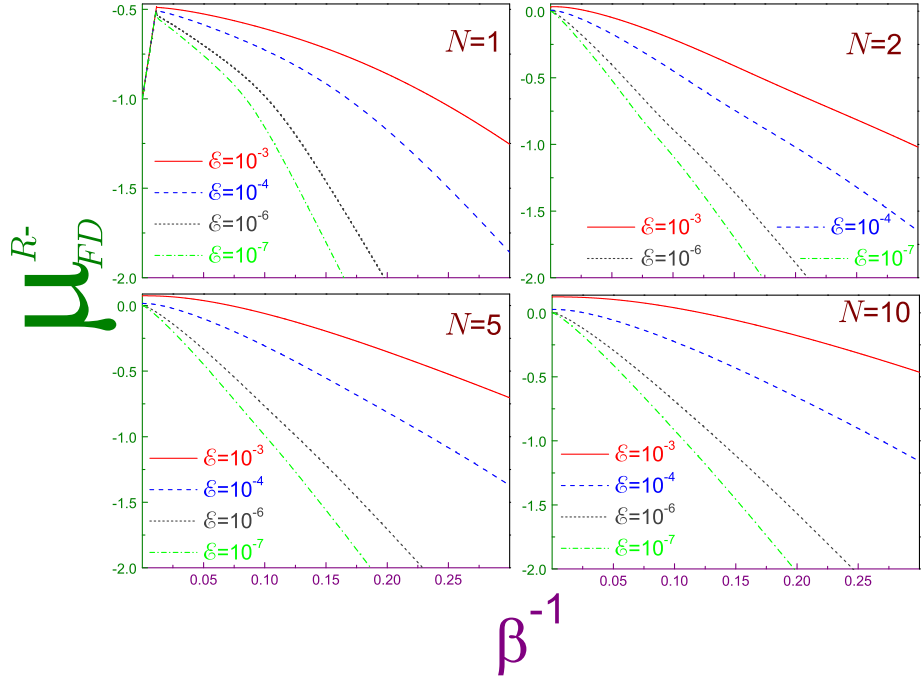}
\caption{\label{FermiFDnegativeLambdaFig1}
FD chemical potential $\mu_{FD}^{R-}$ of the attractive Robin wall as a function of the temperature $\beta^{-1}$ for several numbers $N$ of fermions shown in each panel and different weak electric fields where solid lines are for $\mathscr{E}=10^{-3}$, dashed curves -- for $\mathscr{E}=10^{-4}$, dotted lines -- for $\mathscr{E}=10^{-6}$, and dash-dotted ones are for $\mathscr{E}=10^{-7}$. Note different upper limits of each vertical axis.}
\end{figure}

For the Dirichlet and Neumann wall, utilizing asymptotic expansions of the series
\begin{subequations}\label{AsymptoticSeries2}
\begin{eqnarray}
&&\sum_{n=0}^\infty\frac{1}{be^{t(n+d)^\alpha}\pm1}=\mp\frac{\Gamma(1+1/\alpha)}{t^{1/\alpha}}{\rm Li}_{1/\alpha}\left(\mp b^{-1}\right)-\frac{d-1/2}{b\pm1}\nonumber\\
\label{AsymptoticSeries2_1}
&&+\frac{b}{1+\alpha}\frac{|d-1/2|^{1+\alpha}}{(b\pm1)^2}t-\ldots,\quad t\rightarrow0\\
&&\sum_{n=0}^\infty\frac{(n+d)^\alpha}{be^{t(n+d)^\alpha}\pm1}=\mp\frac{\Gamma(1+1/\alpha)}{\alpha t^{1+1/\alpha}}{\rm Li}_{1+1/\alpha}\left(\mp b^{-1}\right)\nonumber\\
&&-\frac{1}{1+\alpha}\frac{|d-1/2|^{1+\alpha}}{b\pm1}+\frac{b}{1+2\alpha}\frac{|d-1/2|^{1+2\alpha}}{(b\pm1)^2}t\nonumber\\
\label{AsymptoticSeries2_2}
&&-\ldots,\quad t\rightarrow0,
\end{eqnarray}
\end{subequations}
with ${\rm Li}_\alpha(z)$ being a polylogarithm \cite{Lewin1}
$$
{\rm Li}_\alpha(z)=\sum_{k=1}^\infty\frac{z^k}{k^\alpha},
$$
one transforms Eqs.~\eqref{NumberN_1} and \eqref{GrandCanonicalMeanEnergy1} to
\begin{eqnarray}\label{AsymptoteFDchemPoten1}
N&=&\mp\frac{1}{2\pi^{1/2}}\frac{1}{\beta^{3/2}\mathscr{E}}\,{\rm Li}_{3/2}\left(\mp e^{\mu\beta}\right)-\frac{\xi}{e^{-\mu\beta}\pm1}\\
\label{AsymptoteFDenergy1}
\frac{\langle E\rangle}{\mathscr{E}^{2/3}}&=&\mp\frac{3}{4\pi^{1/2}}\frac{1}{\beta^{5/2}\mathscr{E}^{5/3}}\,{\rm Li}_{5/2}\left(\mp e^{\mu\beta}\right)
\end{eqnarray}
valid at $\beta\mathscr{E}^{2/3}\rightarrow0$. Here, $\xi\equiv\xi^{_N^D}=\pm\frac{1}{4}$. Assuming that the chemical potential is large and negative, the following ensemble independent expressions are obtained:
\begin{eqnarray}\label{AsymptoteFDchemPoten2}
\mu^{_N^D}&=&\frac{1}{\beta}\ln\frac{N}{\frac{1}{2\pi^{1/2}}\frac{1}{\beta^{3/2}\mathscr{E}}\mp\frac{1}{4}},\quad\beta\mathscr{E}^{2/3}\rightarrow0\\
\label{AsymptoteFDenergy2}
\left\langle E^{_N^D}\right\rangle&=&N\left(\frac{3}{2\beta}\pm\frac{3}{4}\pi^{1/2}\beta^{1/2}\mathscr{E}\right),\quad\beta\mathscr{E}^{2/3}\rightarrow0.
\end{eqnarray}
Observe that the specific heat per particle $c_N\equiv c/N$ derived from Eq.~\eqref{AsymptoteFDenergy2} exactly coincides with the one from Eq.~\eqref{AsymptoteCanonicalDNsmallY1} for the canonical distribution. Thus, the well-known general result \cite{Dalarsson1} of the independence at the high temperatures of the thermodynamic properties on the ensemble is confirmed again \cite{Olendski3}.

For the attractive Robin wall at the weak fields, Eq.~\eqref{NumberN_1} turns to
\begin{eqnarray}
N&=&\frac{1}{e^{-\beta}e^{-\mu\beta}\pm1}\mp\frac{1}{2\pi^{1/2}}\frac{1}{\beta^{3/2}\mathscr{E}}\,{\rm Li}_{3/2}\left(\mp e^{\mu\beta}\right)\nonumber\\
\label{AsymptoteFDchemPoten3}
&-&\frac{1}{4}\frac{1}{e^{-\mu\beta}\pm1},\quad\mathscr{E}\ll1,\,\beta\mathscr{E}^{2/3}\ll1,
\end{eqnarray}
where also a smallness of the product $\beta\mathscr{E}^{2/3}$ has been assumed. Besides, considering the coefficient $e^{\mu\beta}$ as a tiny parameter too, one gets:
\begin{eqnarray}
N=\frac{1}{e^{-\beta}e^{-\mu\beta}\pm1}+\frac{1}{2\pi^{1/2}}\frac{e^{\mu\beta}}{\beta^{3/2}\mathscr{E}}-\frac{1}{4}\frac{1}{e^{-\mu\beta}\pm1},\nonumber\\
\label{AsymptoteFDchemPoten4}
\mathscr{E}\ll1,\,\beta\mathscr{E}^{2/3}\ll1.
\end{eqnarray}
This equation has a logarithmic solution for the chemical potential $\mu$ but its explicit form is too unwieldy; therefore, to simplify our qualitative analysis, we disregard the last right-hand side term to obtain:
\begin{subequations}\label{AsymptoteFDchemPoten5}
\begin{align}\label{AsymptoteFDchemPoten5_1}
\mu^{_{BE}^{FD}}=\frac{1}{\beta}\ln\frac{1}{2}\frac{r_1^{_{BE}^{FD}}+N\mp1\mp re^{-\beta}}{r},\\
\intertext{where, for convenience, the factors $$r=\frac{1}{2\pi^{1/2}}\frac{1}{\beta^{3/2}\mathscr{E}}$$ and
$$r_1^{_{BE}^{FD}}=\left(r^2e^{-2\beta}+2re^{-\beta}\pm2Nre^{-\beta}+1\mp2N+N^2\right)^{1/2}$$ have been used. For one particle, $N=1$, Eq.~\eqref{AsymptoteFDchemPoten5_1} simplifies to}
\label{AsymptoteFDchemPoten5_2}
\mu_{N=1}^{_{BE}^{FD}}=\frac{1}{\beta}\left\{\begin{array}{c}
\ln\frac{1}{2}e^{-\beta}\left(\sqrt{1+8\pi^{1/2}\beta^{3/2}\mathscr{E}e^\beta}-1\right)\\
\ln\frac{1}{2}\left[e^{-\beta}\left(\sqrt{1+16\pi\beta^3\mathscr{E}^2e^{2\beta}}+1\right)+4\pi^{1/2}\beta^{3/2}\mathscr{E}\right]
\end{array}\right..
\end{align}
\end{subequations}
The corresponding mean energy under the same assumptions reads:
\begin{eqnarray}
\langle E\rangle=-\frac{1}{e^{-\beta}e^{-\mu\beta}\pm1}+\frac{3}{4\pi^{1/2}}\frac{e^{\mu\beta}}{\beta^{5/2}\mathscr{E}},\nonumber\\
\label{AsymptoteFDenergy3}\mathscr{E}\ll1,\,\,\beta\mathscr{E}^{2/3}\ll1.
\end{eqnarray}
The associated heat capacity is calculated from this expression with the help of Eq.~\eqref{HeatCapacity1_1}.

\subsection{Fermions}
\begin{figure}
\centering
\includegraphics[width=\columnwidth]{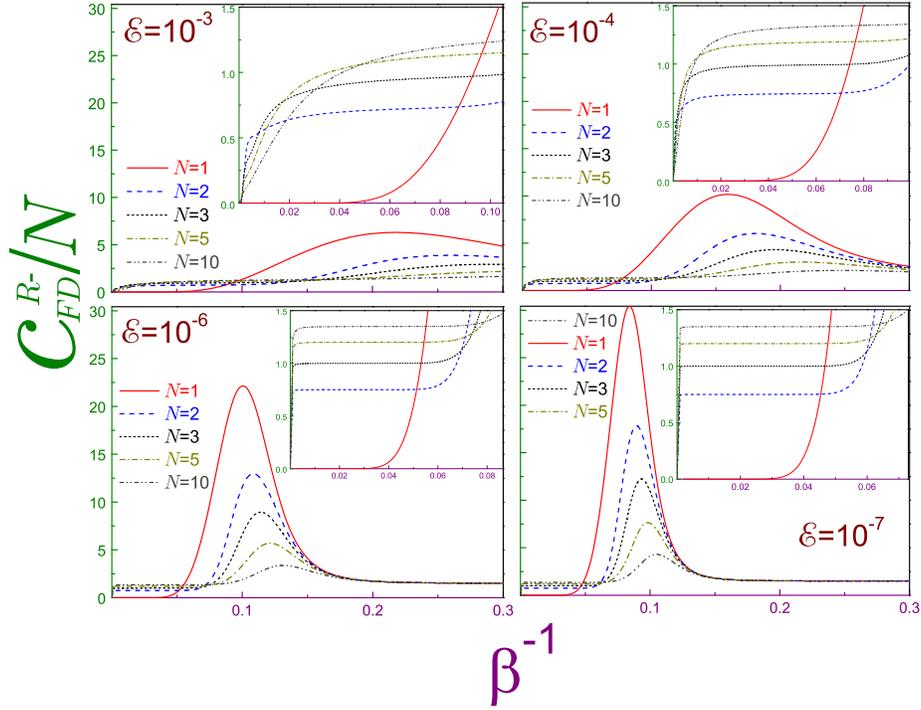}
\caption{\label{HeatCapacityFDnegativeLambdaFig1}
Heat capacity per particle $c_N^{R-}$ of the attractive Robin wall as a function of the temperature $\beta^{-1}$ for different weak electric fields shown in each panel and several numbers $N$ of fermions where solid lines are for $N=1$, dashed curves -- for $N=2$, dotted lines -- for $N=3$, dash-dotted ones -- for $N=5$, and dash-dot-dotted curves are for $N=10$. Insets of each panel show enlarged views at the small temperatures where the formation of the plateaus from Eq.~\eqref{Plateau1} is demonstrated.}
\end{figure}

Consider first the case of the FD distribution. For one fermion, $N=1$, at the attractive Robin wall its specific heat reaches the maximum of
\begin{equation}\label{MaximumHeatCapacity2}
\left.c_{max}^{FD}\right|_{N=1}^{R-}=\frac{\beta_{FDmax}^2}{2^{1/2}(2^{1/2}+1)^2}
\end{equation}
at the inverse temperature
\begin{equation}\label{TemperatureOfMaximumFD1}
\beta_{FDmax}=\frac{3}{2}\,W\!\!\left(\frac{1}{6\pi^{1/3}\mathscr{E}^{2/3}}\right)
\end{equation}
that is a solution of equation
\begin{equation}\label{EquationFD1}
8\pi^{1/2}\mathscr{E}\beta^{3/2}e^\beta=1.
\end{equation}
These expressions show that, similar to the canonical ensemble, the heat capacity unrestrictedly grows as $\ln^2\mathscr{E}$ with the vanishing field.

Fig.~\ref{FermiFDnegativeLambdaFig1} depicts the Fermi energies of the attractive surface as functions of the temperature for several numbers $N$ and weak electric intensities $\mathscr{E}$. It was shown before \cite{Olendski3} that for the structures for which the differences between the several adjacent levels are of the same order, in particular, $\delta E_{N-1}\sim\delta E_N\sim\delta E_{N+1}\sim\ldots$, the fermionic chemical potential $\mu_N$ after its rapid increase from $E_{N-1}$ at $T=0$ exhibits a temperature-independent plateau $(E_{N-1}+E_N)/2$ whose width on the $T$ axis is a function of the differences $\delta E_n$, $n=0,\ldots,N-1,N$. It is a consequence of the interaction at the small temperatures between these two neighboring levels only, which, at $T=0$ were the highest occupied and lowest unoccupied orbitals. Reminiscences of these flat parts can be seen in the figure for the stronger voltages and/or larger number of fermions. The plateaus are completely absent at the extremely weak fields and $N=1$ as the upper left panel of Fig.~\ref{FermiFDnegativeLambdaFig1} demonstrates. Here, the split-off level with the growth of the temperature interacts not only with the lowest unoccupied at $T=0$ state but, in addition to this, with the huge number of its adjacent counterparts. As a result, the Fermi energy has a sharp peak $\mu_1\cong(E_0+E_1)/2\approxeq-1/2$ after which the occupation of the positive-energy levels pushes it downward with the steepness of the curve increasing for the smaller $\mathscr{E}$. At the large temperatures, it is described by the Dirichlet dependence from Eq.~\eqref{AsymptoteFDchemPoten2}.

Corresponding heat capacities are shown in Fig.~\ref{HeatCapacityFDnegativeLambdaFig1}. As discussed above, a specific heat of the attractive Robin interface possesses at the weak fields a conspicuous maximum whose magnitude grows with the decreasing electric intensity $\mathscr{E}$ while its location shifts to the lower $\beta^{-1}$. A physical explanation of this phenomenon is the same as for the canonical ensemble discussed in the previous section. Quantitatively, at the fixed voltage the extremum for one fermion is smaller and located at the higher temperatures, as compared to the canonical distribution, with some data provided in Table~\ref{Table1}.

The increasing number of the electrons in the system subdues the peak and moves it to the warmer temperatures since the higher lying at $T=0$ particles impede the interaction of their lowest counterpart with the levels from the positive part of the spectrum. Mathematically, this can be shown from the asymptotic expansion of the solution of Eq.~\eqref{AsymptoteFDchemPoten4}, which reads for the large $N$:
\begin{equation}\label{AsymptoteFDchemPoten6}
\mu^{_{BE}^{FD}}=\frac{1}{\beta}\left(\ln N-\ln2\pi^{1/2}\beta^{3/2}\mathscr{E}\mp\frac{3}{4}\frac{1}{N}\right),\quad N\gg1.
\end{equation}
Keeping the first two largest terms only leads to the following expression for the heat capacity per particle:
\begin{eqnarray}
c_N^{_{BE}^{FD}}=\frac{3}{2}+\frac{\pi^{1/2}\beta^{5/2}\mathscr{E}(2\beta+3)}{\left(2\pi^{1/2}\beta^{3/2}\mathscr{E}N\pm e^{-\beta}\right)^2}\,e^{-\beta},\nonumber\\
\label{AsymptoteFDheatCap1}\mathscr{E}\ll1,\,\,\beta\mathscr{E}^{2/3}\ll1,\,N\gg1.
\end{eqnarray}
For the case of the fermions, this equation shows that the specific heat rapidly flattens for the larger number of the electrons in the system. This is exemplified in Fig.~\ref{HeatCapacityFDnegativeLambdaFig1} and Table~\ref{Table1}.

Another feature worth mentioning is the fact that the maximum is preceded on the temperature axis by the flat plateau with its value equal to
\begin{equation}\label{Plateau1}
c_{N_{pl}}^{R-}=\frac{3}{2}\frac{N-1}{N}.
\end{equation}
It is explained by the statistical interplay between the fermions; namely, at the zero temperature $N-1$ particles occupy closely located on the $E$ axis positive-energy states, Eq.~\eqref{AsymptoticEnergies2_1}, and the remaining one has the negative energy from Eq.~\eqref{AsymptoticEnergies2_Negative0}. As the density of states at $E>0$ is very large, each of $N-1$ electrons attains the classical heat capacity of $3/2$ at the very small temperature that decreases logarithmically with the field. Further warming of the system does not cause any appreciable change of the specific heat as the lowest level is separated from its positive counterparts by the unit energy gap. Only when the thermal quantum is strong enough to promote the fermion to the positive energies, it does start to contribute to the specific heat. Similar to the canonical distribution, for the smaller electric intensities there are more closely packed positive-energy states in the same energy interval; accordingly, the peak of the heat capacity gets higher for the weaker fields. The formation of the plateaus is clearly seen in the insets of Fig.~\ref{HeatCapacityFDnegativeLambdaFig1}. The widths of these temperature-independent flat parts shrink at the lower voltages.

\newpage
\begin{sidewaystable}
{\tiny
\caption{Peak values of the specific heat per particle ${c_N}_{max}$ and temperatures $\beta_{max}^{-1}$ at which they are achieved for the canonical and several numbers of particles $N$ of the two grand canonical ensembles at some weak electric fields}
\centering
\resizebox{23.5cm}{!}{ 
\begin{tabular}{c||c c||c c|c c|c c|c c||c c|c c|c c|c c|c c|c c}
\hline\hline
\multirow{3}{*}{$\mathscr{E}$}&\multicolumn{2}{c||}{\multirow{2}{*}{Canonical}}&\multicolumn{8}{c||}{Fermions}&\multicolumn{12}{c}{Bosons}\\
\cline{4-23}
&\multicolumn{2}{c||}{}&\multicolumn{2}{c|}{N=1}&\multicolumn{2}{c|}{N=2}&\multicolumn{2}{c|}{N=5}&\multicolumn{2}{c||}{N=10}&\multicolumn{2}{c|}{N=1}&\multicolumn{2}{c|}{N=2}&\multicolumn{2}{c|}{N=5}&\multicolumn{2}{c|}{N=10}&\multicolumn{2}{c|}{N=1000}&\multicolumn{2}{c}{N=100000}\\
\cline{2-23}
&$\beta_{max}^{-1}$&$c_{max}$&$\beta_{max}^{-1}$&$c_{max}$&$\beta_{max}^{-1}$&${c_N}_{max}$&$\beta_{max}^{-1}$&${c_N}_{max}$&$\beta_{max}^{-1}$&${c_N}_{max}$&$\beta_{max}^{-1}$&$c_{max}$&$\beta_{max}^{-1}$&${c_N}_{max}$&$\beta_{max}^{-1}$&${c_N}_{max}$&$\beta_{max}^{-1}$&${c_N}_{max}$&$\beta_{max}^{-1}$&${c_N}_{max}$&$\beta_{max}^{-1}$&${c_N}_{max}$\\
\hline
$10^{-3}$&0.2504&7.815&0.2180&6.295&0.2605&3.874&0.3350&2.219&0.4080&1.766&0.2812&8.124&0.3052&8.199&0.3573&8.115&0.4179&7.828&2.3157&3.934&30.800&2.361\\
$10^{-4}$&0.1750&13.154&0.1594&10.164&0.1807&6.028&0.2163&3.006&0.2483&2.122&0.1907&14.011&0.2024&14.369&0.2272&14.600&0.2545&14.391&0.8915&6.922&7.9544&2.990\\
$10^{-5}$&0.1324&20.538&0.1240&15.416&0.1364&9.035&0.1566&4.153&0.1731&2.650&0.1413&22.327&0.1481&23.216&0.1619&24.223&0.1765&24.463&0.4399&13.192&2.4146&4.447\\
$10^{-6}$&0.1056&30.093&0.1005&22.145&0.1086&12.957&0.1212&5.694&0.1314&3.375&0.1113&33.240&0.1155&34.950&0.1240&37.249&0.1328&38.400&0.2644&24.505&0.9228&7.786\\
$10^{-7}$&0.0873&41.910&0.0840&30.417&0.0895&17.837&0.0984&7.654&0.1051&4.312&0.0912&46.875&0.0940&49.694&0.0997&53.847&0.1056&56.414&0.1815&42.095&0.4513&14.828\\
[1ex]
\hline
\end{tabular}
}
\label{Table1}
}
\end{sidewaystable}

\subsection{Bosons}
\begin{figure}
\centering
\includegraphics[width=\columnwidth]{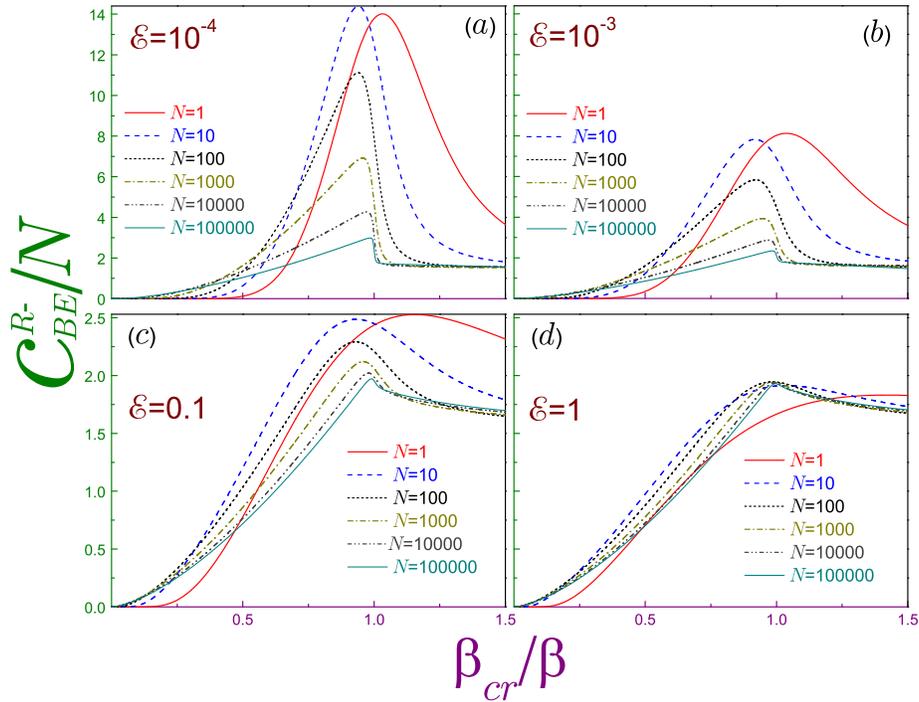}
\caption{\label{BoseHeatCapacityNegLambdaFig1}
Bosonic heat capacity per particle $c_N^{R-}$ of the attractive Robin wall as a function of the normalized temperature $\beta_{cr}/\beta$ for (a) $\mathscr{E}=10^{-4}$, (b) $\mathscr{E}=10^{-3}$, (c) $\mathscr{E}=0.1$ and (d) $\mathscr{E}=1$ and several numbers $N$ of bosons where thick solid lines are for $N=1$, dashed curves -- for $N=10$, dotted lines -- for $N=100$, dash-dotted curves -- for $N=1000$, dash-dot-dotted ones are for $N=10000$, and thin solid lines -- for $N=100000$. Note different upper limits for the lower and upper panels.}
\end{figure}

A remarkable property of the collective behavior of this type of particles is a formation of the condensate when the overwhelming majority of bosons reside in the ground level.  Impressive experimental discovery \cite{Anderson1,Bradley1,Davis1} of this state of matter predicted in 1925 \cite{Bose1} (the history of its theoretical derivation and subsequent development is described in, e.g., Ref.~\cite{Gaponenko1}) reignited a huge interest in the topic \cite{Dalfovo1,Pitaevskii1,Leggett1,Pethick1}. Relevant to our research, let us mention a theoretical analysis of the influence of the different forms of the confining potential \cite{Dalfovo1,Ketterle1,Druten1,Napolitano1,Mullin1,Bagnato1,Bagnato2} and/or BC \cite{Olendski3,Krueger1,Sonin1,Greenspoon1,Zasada1,Goble1,Goble2,Goble3,Grossmann1,Pathria1,Greenspoon2,Barber1,Hasan1} on the properties of the ideal Bose gas. A quantitative measure of the BE condensation is provided by the ground state occupation $n_0(\mathscr{E};\beta)$, i.e., a ratio of the number of bosons $N_0$ in the ground state
\begin{equation}\label{Number_N0}
N_0=\frac{1}{e^{(E_0-\mu)\beta}-1}
\end{equation}
to the total number of corpuscles in the system:
\begin{equation}\label{Number_n0}
n_0=\frac{N_0}{N}.
\end{equation}
As this fraction is a function of $\beta^{-1}$, we, following Ketterle and van Druten \cite{Ketterle1}, define and  will use the critical temperature $\beta_{cr}^{-1}$ as the largest temperature at which the BE condensation can still be observed what describes the configuration with the chemical potential equal to the ground-state energy, $\mu=E_0$, and with no particles dwelling in the lowest level, $N_0=0$:
\begin{equation}\label{CriticalTemp1}
\sum_{n=1}^\infty\frac{1}{e^{(E_n-E_0)\beta_{cr}}-1}=N.
\end{equation}
Critical temperature is an increasing function of the number of bosons in the system $N$ and electric field $\mathscr{E}$; for example, for its small values, $T_{cr}^{R-}\ll1$, one has
\begin{equation}\label{CriticalTemp2}
\beta_{cr}^{R-}=\frac{3}{2}\,{\rm W}\!\left(\frac{4^{2/3}}{6}\frac{1}{\pi^{1/3}N^{2/3}\mathscr{E}^{2/3}}\right),\quad\mathscr{E}\ll1.
\end{equation}
\begin{figure}
\centering
\includegraphics[width=\columnwidth]{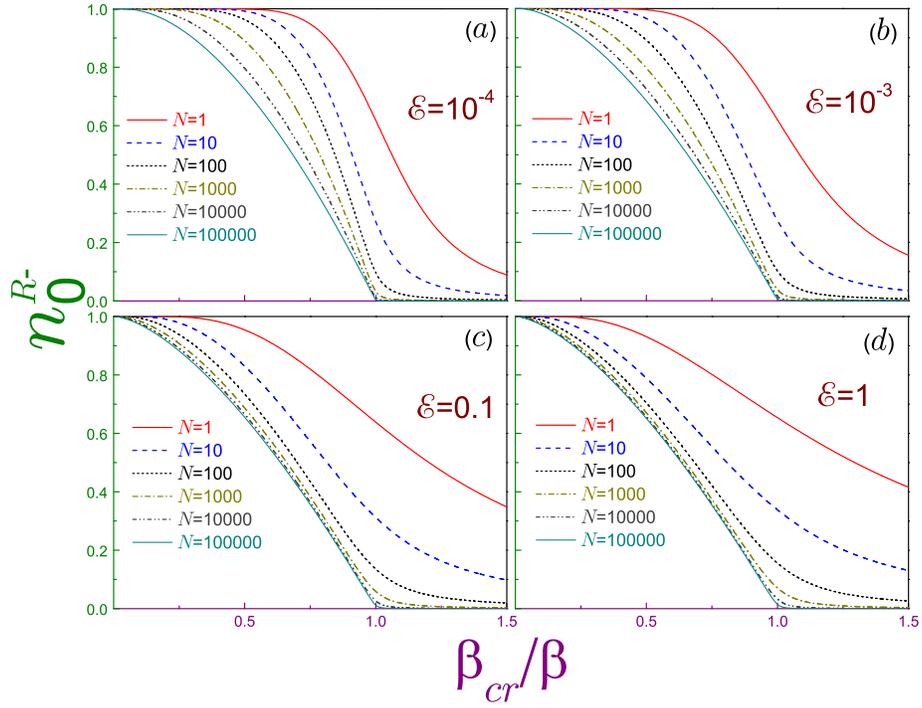}
\caption{\label{BoseGroundPopulNegLambdaFig1}
Ground-state occupation $n_0^{R-}$ of the attractive Robin wall as a function of the normalized temperature. The same conventions as in Fig.~\ref{BoseHeatCapacityNegLambdaFig1} are used.}
\end{figure}
\begin{figure}
\centering
\includegraphics[width=\columnwidth]{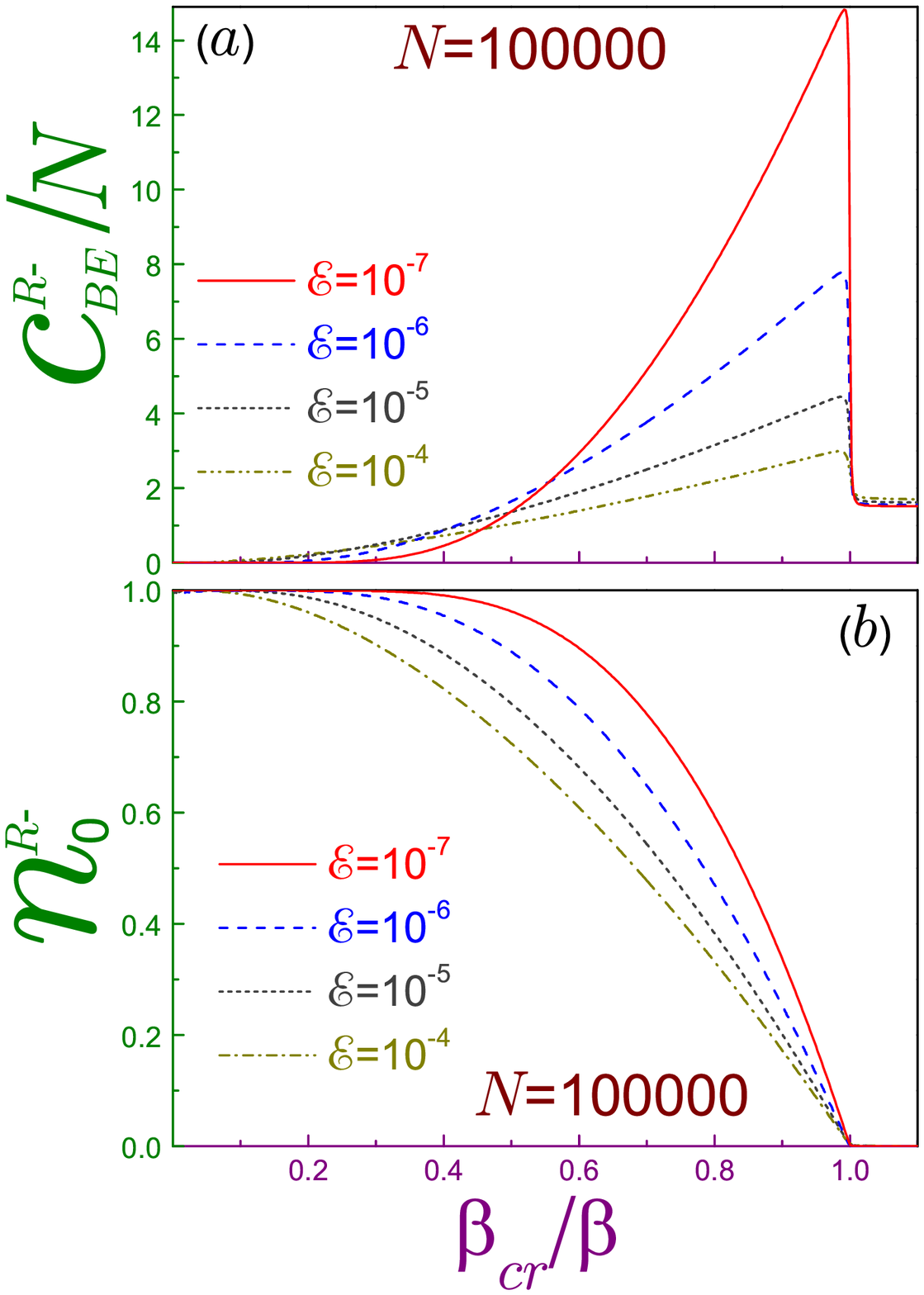}
\caption{\label{BoseNegLambdaFig1}
Bosonic (a) heat capacity per particle $c_N^{R-}$ and (b) ground-state occupation $n_0^{R-}$ of the attractive Robin surface as a function of the normalized temperature for several electric fields and fixed number of bosons $N=10^5$. Solid lines denote dependencies for $\mathscr{E}=10^{-7}$, dashed curves -- for $\mathscr{E}=10^{-6}$, dotted ones - for $\mathscr{E}=10^{-5}$, and dash-dotted lines are for $\mathscr{E}=10^{-4}$.}
\end{figure}

\begin{figure}
\centering
\includegraphics[width=\columnwidth]{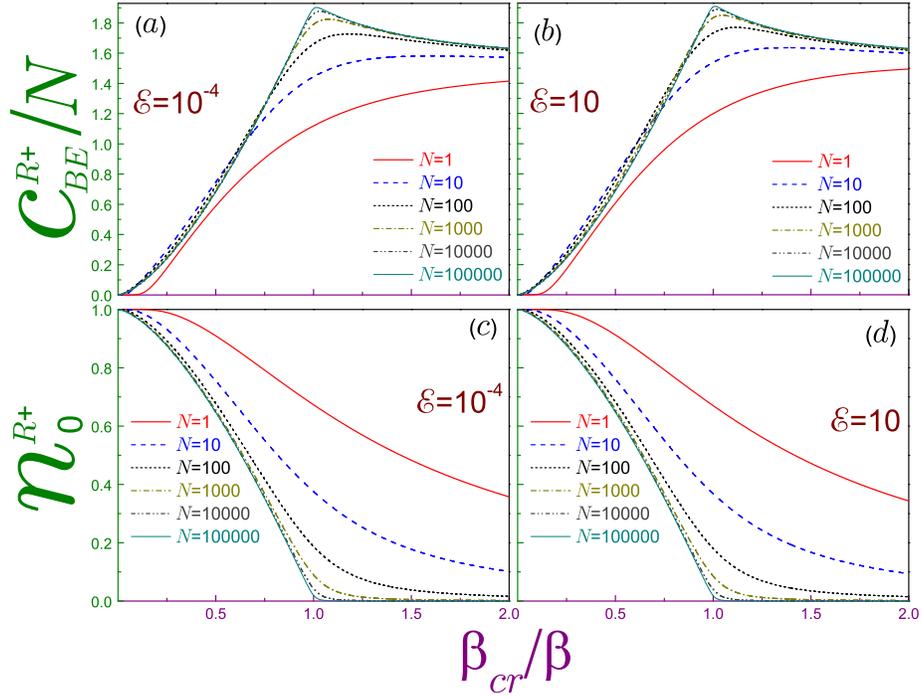}
\caption{\label{BosePosLambdaFig1}
Upper (lower) panels show bosonic heat capacities per particle $c_N^{R+}$ (ground-state occupations $n_0^{R+}$) of the repulsive Robin surface as a function of the scaled temperature where left subplots [(a) and (c)] are for $\mathscr{E}=10^{-4}$ while the right ones [(b) and (d)] correspond to $\mathscr{E}=10$. The same line convention as in Fig.~\ref{BoseHeatCapacityNegLambdaFig1} is used.}
\end{figure}

Fig.~\ref{BoseHeatCapacityNegLambdaFig1} draws BE heat capacity per particle of the attractive surface as a function of temperature (in units of its critical counterpart) for several fields and numbers of bosons. It exemplifies that for the one particle  the $c_N$ behavior is qualitatively the same as for the fermion; namely, as the applied voltage lowers, the extremum of the specific heat grows and is shifted to the colder temperatures. Some field-specific values provided in Table~\ref{Table1} show that the bosonic peak is the largest among the three ensembles and is achieved at the highest $T$. Note that, depending on the electric intensity, with the growth of the moderate $N$ the maximum can increase, as panel (a)  and data from the Table demonstrate. As the number of bosons becomes sufficiently large, the domination of the statistical effects starts to reveal itself leading to the formation from the resonance of an asymmetric cusp-like structure with its right side leaning more vertically with the increasing $N$. Observe that this rapid descent of the heat capacity occurs at the temperature closer and closer coinciding with the critical one. Simultaneously, the ground-state population becomes a steeper function of the temperature almost turning to zero at $\beta_{cr}^{-1}$, as Fig.~\ref{BoseGroundPopulNegLambdaFig1} depicts. At zero temperature, all bosons stay on the ground level, $\left. n_0\right|_{T=0}=1$. The heating of the wall pushes them away from the lowest state with the decrease of $n_0$ at  $T\ll T_{cr}$ ($T\sim T_{cr}$) being flatter (more  precipitous) at the smaller fields. For the infinite number of bosons, $N=\infty$, the ground orbital will be completely depopulated at $T\geq T_{cr}$ what means a full destruction of the BE condensate by the warming of the structure but due to the finiteness of $N$ a tiny field-dependent fraction $n_0$ persists for the temperatures above the critical one. To emphasize the role of the varying field in the process of BE condensate evolution and destruction, Fig.~\ref{BoseNegLambdaFig1} depicts heat capacity per particle and ground-state occupation at the fixed number of bosons $N=100000$ and several very low voltages. It is seen that the decreasing electric intensity causes, similar to the canonical and FD ensembles, an increase of the peak value of the heat capacity with the lowest level depopulation taking place at the higher normalized temperatures $T/T_{cr}$. In the limit of the vanishing field the ground-state occupation turns into the step function $h(z)=\left\{\begin{array}{cc}
1,&z\geq0\\
0,&z<0
\end{array}\right.$:
\begin{subequations}\label{AsymptoteBose1}
\begin{align}\label{AsymptoteBose1_n0}
n_0^{R-}(\beta)&\xrightarrow[\mathscr{E}\rightarrow0]{}h(\beta_{cr}/\beta),
\intertext{while the critical temperature itself approaches the absolute zero, as it follows from Eq.~\eqref{CriticalTemp2}:}
\label{AsymptoteBose1_Tcr}
T_{cr}^{R-}&\xrightarrow[\mathscr{E}\rightarrow0]{}0.
\end{align}
\end{subequations}
The expression for $\beta_{cr}$ from Eq.~\eqref{CriticalTemp2}, which was derived directly from Eq.~\eqref{CriticalTemp1} under the assumptions $\beta_{cr}\gg1$ and $\mathscr{E}\ll1$, is obtained also by zeroing the denominator in Eq.~\eqref{AsymptoteFDheatCap1} that, in addition to the above two conditions, is valid for the large $N$. The infinite heat capacity at the critical temperature that follows from the latter formula is due to the neglect of the higher-order terms, which, if included into the consideration, smooth out the heat capacity peak to the finite field-dependent magnitude, as exact calculations depicted in Fig.~\ref{BoseNegLambdaFig1}(a) demonstrate. The cusp-like structure of the heat capacity that was predicted theoretically for a number of the confining potentials \cite{Druten1,Napolitano1,Grossmann2,Haugerud1,Haugset1,Goswami1} and observed experimentally for, e.g., the dilute Bose gas of $^{87}{\rm Rb}$ atoms \cite{Ensher1}, for the infinite number of particles, $N=\infty$, turns at $T=T_{cr}$ into the discontinuity that is a manifestation of the phase transition; in our case, it is a transition from the BE condensate to the normal phase of the noninteracting particles in the linear potential. This justifies the definition of the critical temperature from Eq.~\eqref{CriticalTemp1}. There has been a lively discussion of the influence of the shape and dimensionality of the confining potential on the existence, form and evolution of this feature \cite{Olendski3,Pitaevskii1,Pethick1,Druten1,Napolitano1,Grossmann2,Haugerud1,Haugset1,Goswami1}. As our results show, the sharpness of the resonance can be effectively controlled by the voltage applied to the attractive Robin surface. For comparison, Fig.~\ref{BosePosLambdaFig1} shows the heat capacity and ground-level occupation of the repulsive wall at the quite strong and very weak fields. It is seen that in terms of the scaled temperature $T/T_{cr}$ the shape of the curves is practically independent of the electric intensity. The increasing number of bosons leads to the formation of the maximum of the specific heat at $\beta^{-1}\sim\beta_{cr}^{-1}$ but, contrary to the attractive surface, this extremum does not grow to infinity rather saturating to the value lying below 2. Contrasting the results of the negative and positive Robin walls shows once again the crucial role of the split off level in the dramatic evolution of the thermodynamic properties.

\section{Concluding Remarks}\label{Conclusions}
A theoretical analysis of the thermodynamic properties of the Robin wall in the perpendicular electric field $\mathscr{E}$ discovered that the heat capacity of the attractive surface for the canonical and either type of the grand canonical ensemble exhibits a pronounced maximum with its peak increasing at the vanishing voltage as $\ln^2\mathscr{E}$ while the corresponding temperature goes to zero as $-1/\ln\mathscr{E}$. This phenomenon that is absent for the repulsive interface is explained by the peculiarities of the associated energy spectrum; namely, at $\mathscr{E}\rightarrow0$ it is characterized by the quasi continuum of the positive-energy states and, additionally, the split off level with $E<0$ \cite{Olendski1}. At the cold temperatures, the heating of structure enforces the transitions from this orbital to a huge number of the higher lying states leading in this way to the giant enhancement of the specific heat $c_V$. Increasing number $N$ of the fermions decreases the extremum while for the BE distribution for the quite large $N$ the resonance changes into the cusp-like structure, which is a manifestation of the phase transition into the condensate regime with the critical temperature being strongly $\mathscr{E}$- and $N$-dependent. Thus, the applied voltage allows to effectively control the thermodynamics of the Robin quantum wall.

Having seen the dramatic evolution of the thermodynamic properties predicted above, one might wonder about their experimental verification. In the previous analysis of the attractive field-free Robin wall \cite{Seba1,Pazma1,Fulop1,Belchev1,Georgiou1} it was conjectured that this model approximates the piecewise continuous potentials that can be grown by the modern semiconductor technologies. Consider, for example,  the term $V(x)$ in Eq.~\eqref{Schrodinger1} of the form (below we return to the regular units):
\begin{equation}\label{Potential2}
V(x)=\left\{\begin{array}{cc}
0,&-\infty<x\leq-x_0\\
-V_0,&-x_0<x<0\\
\infty,&x\geq0
\end{array}\right.
\end{equation}
with positive $x_0$ and $V_0$. It is elementary to show that, in addition to its positive spectrum, this asymmetric QW can support bound states with their number depending on the interrelation between $V_0$ and $x_0$; for example, each new localized level $n$ emerges from the positive continuum at the following threshold magnitude of $V_0$:
\begin{equation}\label{Potential3}
V_n^{TH}=\left(n-\frac{1}{2}\right)^2\frac{\pi^2\hbar^2}{2mx_0^2},\quad n=1,2,\ldots.
\end{equation}
It means that for $V_0$ satisfying the condition
\begin{equation}\label{Potential4}
\frac{1}{8}\frac{\pi^2\hbar^2}{mx_0^2}<V_0<\frac{9}{8}\frac{\pi^2\hbar^2}{mx_0^2},
\end{equation}
only one negative-energy state does exist. As a result, this potential forms zero-field energy spectrum complying with all the requirements of the present research. Accordingly, applying to it appropriately directed voltage, it is possible to check the heat capacity enhancement discussed above.

\end{document}